\ifdefined\XeTeXversion\else\ifdefined\directlua\else\pdfoutput=1\fi\fi %
\documentclass[11pt]{article}
\usepackage{thefinai}

\newcommand{\singlefigwidth}{0.62\textwidth} %
\newcommand{\preprintteaser}{}  %
\titleformat{\paragraph}[runin]{\bfseries\color{finaiink}}{}{0pt}{} %

\makeatletter
\newcommand{\partnerlogo}[2]{\includegraphics[height=#1]{logos/#2}}
\newcommand{\finai@logorow}{%
  \resizebox{\linewidth}{!}{%
  \thefinaiwordmark[1.55em]\hspace{1.4em}%
  \partnerlogo{12.5pt}{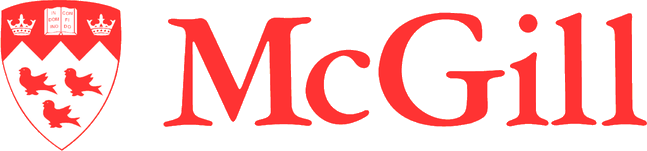}\hspace{1.4em}%
  \partnerlogo{16pt}{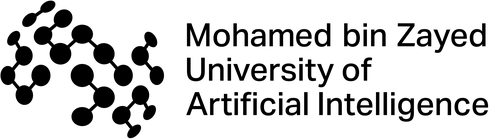}\hspace{1.4em}%
  \partnerlogo{10.5pt}{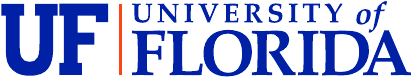}\hspace{1.4em}%
  \partnerlogo{23pt}{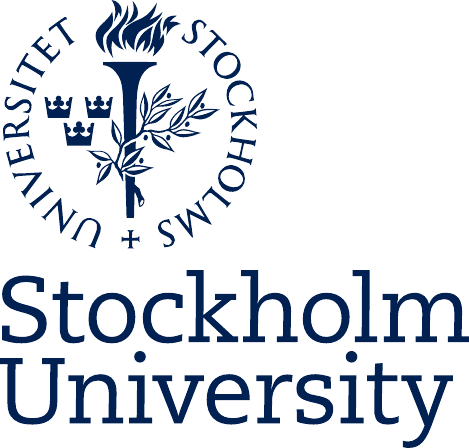}\hspace{1.4em}%
  \partnerlogo{15pt}{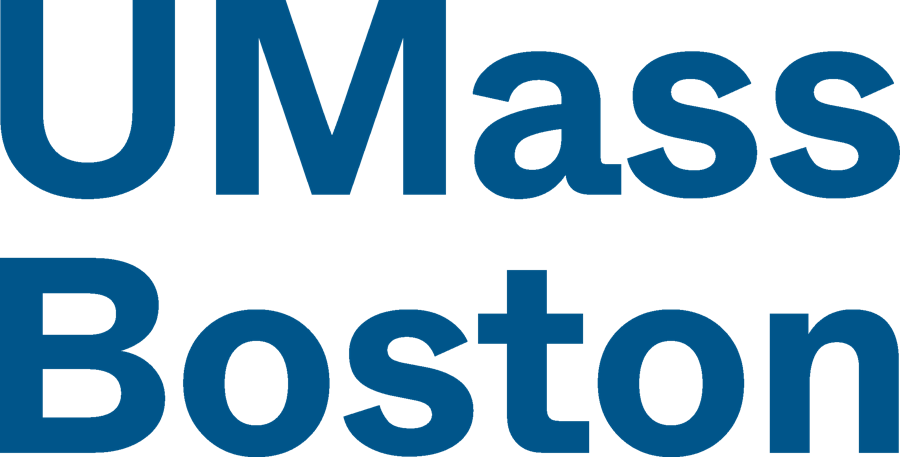}\hspace{1.4em}%
  \partnerlogo{17pt}{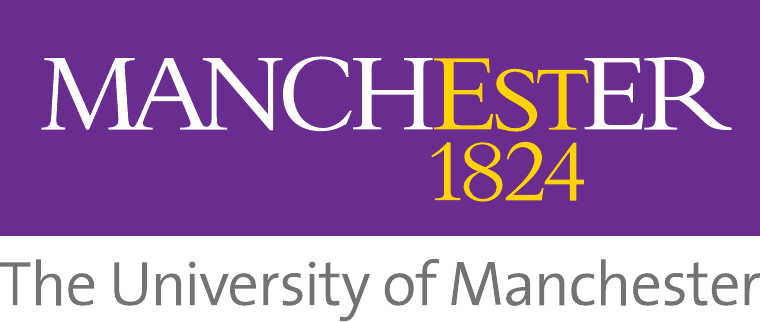}\hspace{1.4em}%
  \partnerlogo{12.5pt}{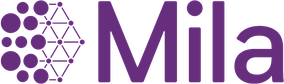}\hspace{1.4em}%
  \partnerlogo{15pt}{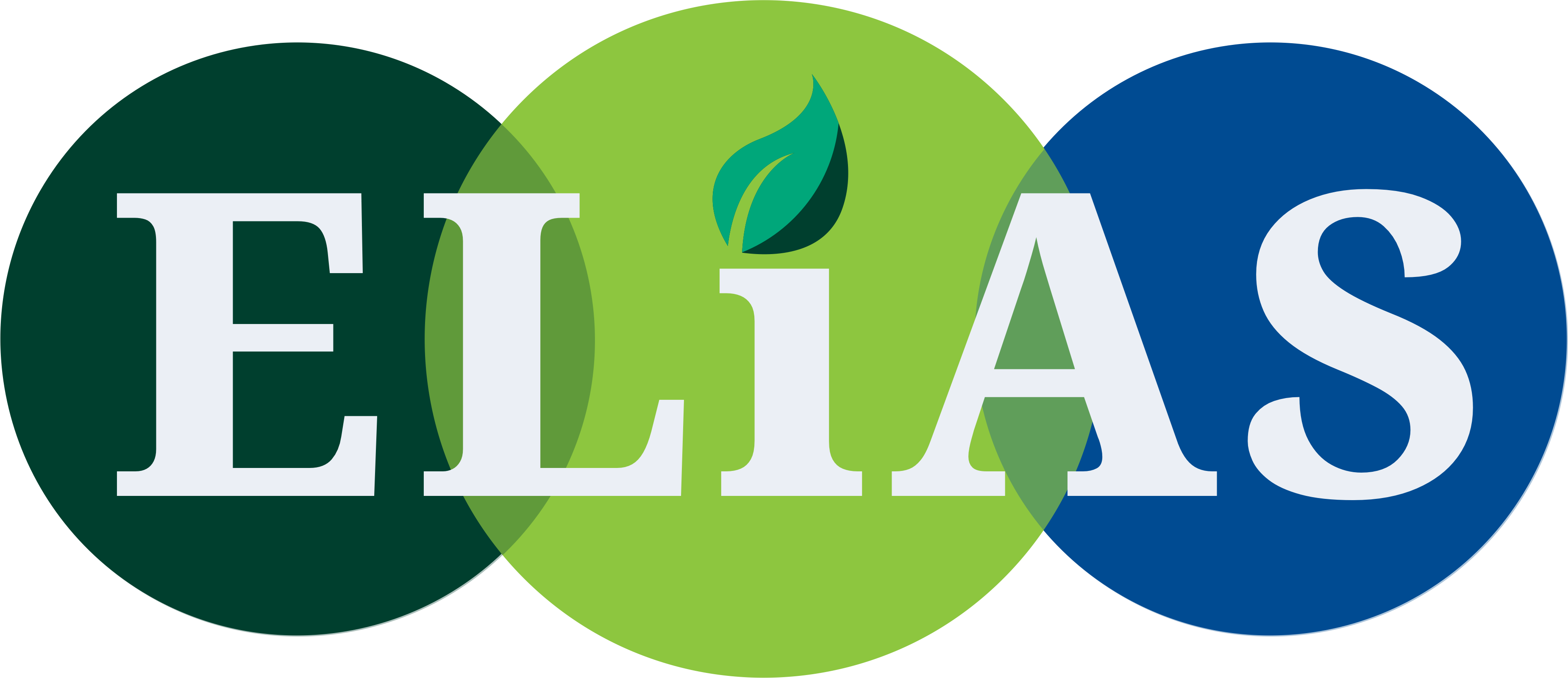}\hspace{1.4em}%
  \partnerlogo{13pt}{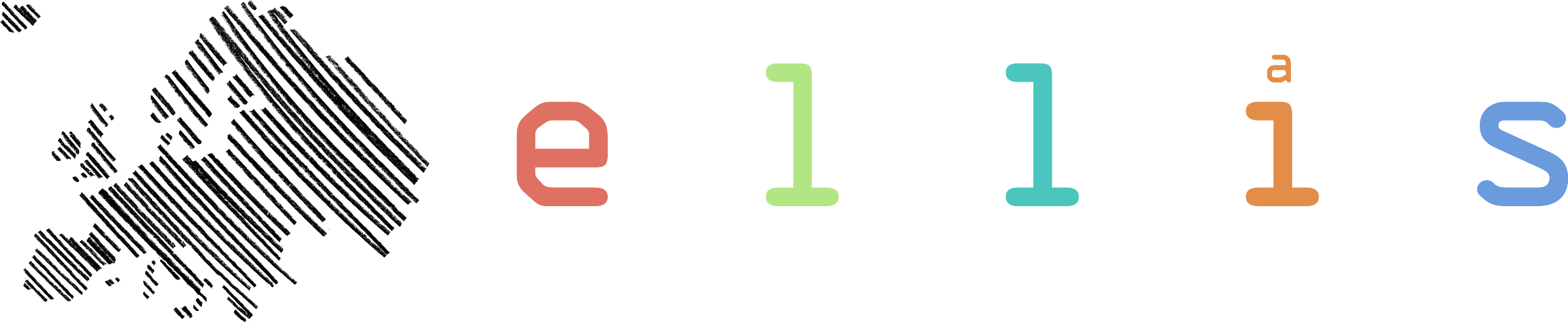}}}
\renewcommand{\maketitle}{%
  \thispagestyle{finaiplain}%
  \begin{tcolorbox}[enhanced,colback=finaicard,frame hidden,boxrule=0pt,arc=7pt,
      left=20pt,right=20pt,top=18pt,bottom=14pt,before skip=0pt,after skip=16pt]
    \raggedright
    \ifx\finai@label\@empty\else{\sffamily\bfseries\scriptsize\color{finaiblue}\MakeUppercase{\finai@label}}\par\vspace{5pt}\fi
    {\sffamily\bfseries\fontsize{18.5}{23}\selectfont\color{finaiblue}\@title\par}%
    \vspace{12pt}%
    {\sffamily\fontsize{10.5}{14}\selectfont\color{finaiink}\finai@authors\par}\vspace{5pt}%
    {\sffamily\footnotesize\color{finaisoft}\finai@affils\par}%
    \ifx\finai@notes\@empty\else\vspace{2pt}{\sffamily\footnotesize\color{finaisoft}\finai@notes\par}\fi
    \ifx\finai@abstract\@empty\else
      \vspace{12pt}{\fontsize{10}{13.2}\selectfont\justifying\finai@abstract\par}%
    \fi
    \vspace{10pt}%
    {\fontsize{9}{12.5}\selectfont\setlength{\parskip}{1.5pt}\finai@metablock}%
    \vspace{8pt}{\color{finaisoft!35}\rule{\linewidth}{0.4pt}}\par\vspace{9pt}%
    \noindent\mbox{}\finai@logorow\mbox{}\par
  \end{tcolorbox}%
  \global\finai@titledonetrue
}
\makeatother

\title{The Market Is the Reward: \\Design Markets, Not Just Traders\\for Human--AI Collective Intelligence}
\date{November 2026}

\correspondence{Xueqing Peng, \href{mailto:xueqing.peng@yale.edu}{xueqing.peng@yale.edu}; Jimin Huang, \href{mailto:jimin.huang@postgrad.manchester.ac.uk}{jimin.huang@postgrad.manchester.ac.uk}}
\author{%
  Yankai Chen\textsuperscript{1,2}, Xueqing Peng\textsuperscript{3,\dag}, Alejandro Lopez-Lira\textsuperscript{4},
  Yuehua Tang\textsuperscript{4}, Micha{\l}~Dzieli{\'n}ski\textsuperscript{5}, Yijia Zhao\textsuperscript{6},
  Jimin Huang\textsuperscript{3,7,\dag}, Xue Liu\textsuperscript{1,2,8}, Sophia Ananiadou\textsuperscript{7,9,10}%
}
\affiliation[1]{McGill University}
\affiliation[2]{Mohamed bin Zayed University of Artificial Intelligence}
\affiliation[3]{The Fin AI}
\affiliation[4]{University of Florida}
\affiliation[5]{Stockholm Business School, Stockholm University}
\affiliation[6]{University of Massachusetts Boston}
\affiliation[7]{University of Manchester}
\affiliation[8]{Mila -- Quebec AI Institute}
\affiliation[9]{ELIAS Manchester Node}
\affiliation[10]{ELLIS Manchester}
\authornote{\textsuperscript{\dag}Corresponding authors}

\begin{document}

\begin{abstract}
Learning agents are becoming market participants, yet machine learning still
treats the market as a fixed environment, although its payoff rule is the
reward agents learn from. This position paper argues that financial AI should
build HyperMarkets for human--AI collective intelligence, and not only the
agents that trade in them: markets whose liquidity provision, information
production, outcome determination, and trading rule are performed or adapted
by learning agents and designed jointly toward explicit market-level
objectives. In a pre-registered simulation with Claude agents trading for
principals, joint design raises a market-level objective by 79\% over separately
tuned functions, through a trading rule that helps only when slow
traders are present, and how evidence is paid for decides whether a stronger
agent earns more for its principal. We state when fixed rules or non-market
aggregation should be preferred and what findings would refute the case.
\end{abstract}

\maketitle

\section{Introduction}
\label{sec:intro}
\ifdefined\preprintteaser
\begin{figure}[!h]
\centering
\includegraphics[width=\textwidth]{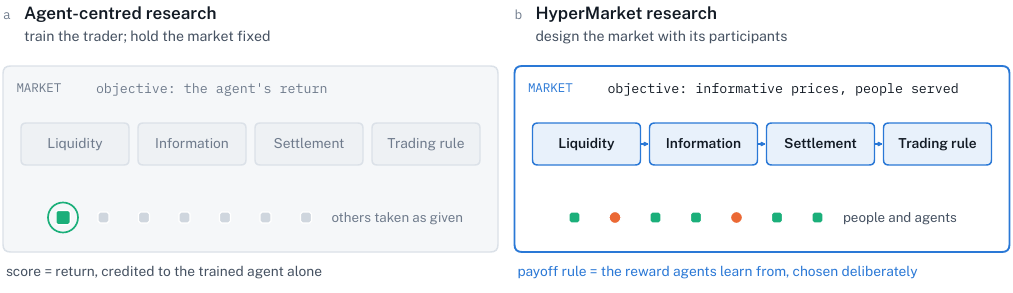}
\caption{\textbf{What is held fixed, and what is designed.} Left: a common
agent-development setup, in which one trained agent (ringed square) acts in a market
whose functions, population, and objective are given, and its return is credited
to it alone. Right: HyperMarket research treats the market's functions as designed
or learned (blue), serving people (orange circles) and agents (green squares) toward a
declared market-level objective. Grey marks what is held fixed.}
\label{fig:shift}
\end{figure}
\else
\begin{figure*}[t]
\centering
\includegraphics[width=\textwidth]{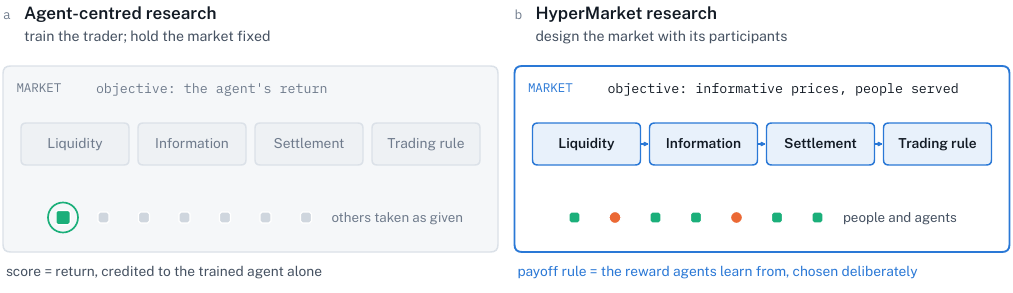}
\caption{\textbf{What is held fixed, and what is designed.} Left: a common
agent-development setup, in which one trained agent (ringed square) acts in a market
whose functions, population, and objective are given, and its return is credited
to it alone. Right: HyperMarket research treats the market's functions as designed
or learned (blue), serving people (orange circles) and agents (green squares) toward a
declared market-level objective. Grey marks what is held fixed.}
\label{fig:shift}
\end{figure*}
\fi

\begin{figure*}[t]
\centering
\includegraphics[width=\textwidth]{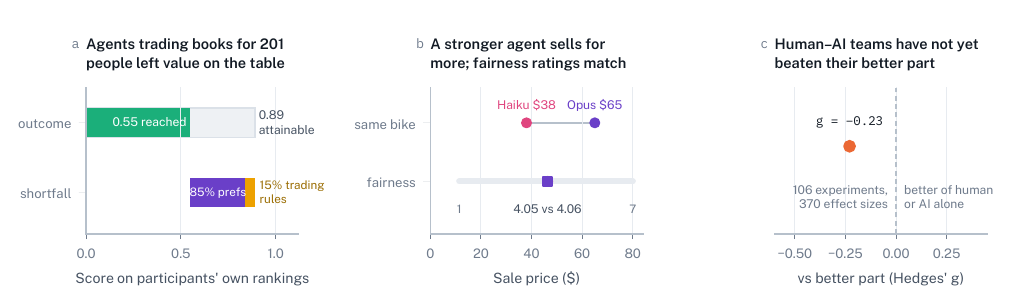}
\caption{\textbf{Agents already trade for people, the market decides what they
achieve, and complementarity is unproven.} (a) In Project Swap, agents traded
books for 201 people under fixed rules; 85\% of the shortfall came from
representing preferences \citep{anthropic2026swap}. (b) In Project Deal, the
same bike sold for \$38 by Haiku and \$65 by Opus, while participants rated
the fairness of the two models' deals 4.06 and 4.05 on a 1--7 scale
\citep{anthropic2026deal}. (c) Across 106 experiments, human--AI combinations
performed worse on average than the better of humans or AI alone
\citep{vaccaro2024combinations}.}
\label{fig:evidence}
\end{figure*}

Learning agents are becoming market participants. Agents built on foundation
models read filings, use tools, and hold positions \citep{yu2024finmem,zhang2024finagent,xiao2024tradingagents,yu2024fincon}, and reinforcement learning now fine-tunes a language model directly on trading rewards \citep{xiong2025flagtrader}. In forecasting, large language models can extract return-relevant information from financial news and predict subsequent stock returns \citep{lopezlira2026chatgpt}, a retrieval-augmented language-model system approaches the accuracy of
aggregated competitive human forecasters \citep{halawi2024forecasting}, and
in a three-month tournament an ensemble of 12 language models was
statistically indistinguishable from a crowd of 925 human forecasters
\citep{schoenegger2024wisdom}. Agents already trade on people's behalf, and what they achieve depends on
the market and on the agent (Figure~\ref{fig:evidence}a,b;
\citealp{anthropic2026deal,anthropic2026swap}). Machine learning has approached these agents as
it approaches most agents: train or prompt a policy, place it in an
environment taken as given, and report its score \citep{liu2021finrl,byrd2020abides}. Financial benchmarks have grown from static tasks \citep{xie2023pixiu,xie2024finben,peng2025multifinben} to agentic workflows \citep{li2024investorbench,cao2026fintrace,peng2026herculean} and to live trading across markets \citep{qian2025agentstrade}, yet each takes the market as the fixed backdrop against which an agent is scored. In a market, this setup leaves out most of
what determines the score.

A market is an unusual environment for a learner, because its dynamics are
produced by other learners and by a rule. Q-learning pricing algorithms in a
repeated oligopoly consistently learn supracompetitive prices without communicating \citep{calvano2020collusion}, learning algorithms with bounded rationality can sustain any feasible and individually rational payoff \citep{cartea2026folk}, and reasoning language-model agents
tacitly collude in pricing games even when instructed otherwise
\citep{collusion2026certification}. When many decision-makers converge on one
algorithm, overall decision quality can fall even if that algorithm is more
accurate for each of them alone \citep{kleinberg2021monoculture}, and models
that share components can homogenize outcomes across deployments
\citep{bommasani2022picking}. A market's payoff rule is also the reward from
which agents trained in it learn: in continuous limit order books, price
correlations break down at millisecond horizons, and the resulting mechanical
arbitrage is something competition only makes faster, so what the market pays
for is speed \citep{budish2015hft}. Holding the market fixed therefore fixes,
without discussion, both the environment and the reward for a growing class of
agents. A return obtained in such a market belongs to the agent and the market
together, yet it is credited to the agent.

What has changed is that the rest of the market can now be learned as well. A
market runs on a few functions: liquidity provision, the production of
information that prices respond to, the determination of how claims resolve,
and the rule that matches orders. Market design has engineered these functions
around human participants \citep{roth2002engineer}: people whose attention is
scarce, whose strategizing is costly, and whose signals are trusted because
they are costly to fake. Each function is now a learning problem. Liquidity
provision is a sequential decision problem addressed with online learning and
reinforcement learning \citep{abernethy2013efficient,spooner2018market}.
Information production is forecasting and generation
\citep{halawi2024forecasting}, and in an online labor market generative AI has
already weakened the signaling value of written applications
\citep{galdin2025talk}. Outcome
determination is evidence assessment; it is still done by people and remains
open \citep{freeman2017crowdsourced}. The rule itself can be learned, as in
learned auctions and economic policies trained against adaptive agents
\citep{duetting2019optimal,zheng2022economist}. For the first time, every
function of a market is something machine learning can build, and all of them
act on one shared environment.

\textbf{We therefore argue that financial AI should build HyperMarkets for
human--AI collective intelligence, and not only the agents that trade in
them.} A HyperMarket is a market in which one or more core functions are
performed or adapted by learning agents, designed jointly with its human and
AI participants toward explicit market-level objectives
(Figure~\ref{fig:shift}).   Complementarity between people and agents is likewise something to
demonstrate (Figure~\ref{fig:evidence}c).

The shift changes the research problem in three ways. The object becomes a
system of interacting learners and a learnable rule: a multi-agent, bilevel
problem rather than a single policy. The objective moves to the level of the
market, such as how informative its prices are and whether the people trading
in it are served rather than exploited; a misspecified market-level objective
can be gamed as surely as any reward \citep{skalse2022defining}. Evaluation
changes with them: a result should report the market in which it was obtained,
including the mechanism, the population, and shared base models. Prediction
markets are our starting case because their prices are public and their claims
resolve, which makes collective intelligence measurable against the world; each
function also has a counterpart in securities markets.

This paper makes four contributions. It reframes the market, for a learning
agent, as both environment and reward, so that market design becomes part of
the learning problem. It gives a function-by-function account of what learning
changes in liquidity provision, information production, outcome
determination, and the trading rule, each tied to existing evidence. It states
falsifiable criteria, grounded in a formal statement of the design problem as a
bilevel program, naming the comparisons that would support the position and
the findings that would weaken it, and runs one such comparison: a
pre-registered simulation with Claude agents, in which joint design pays
through an interaction we did not predict while the channel we did predict
fails. And it proposes concrete actions for
researchers, benchmark builders, venues, and operators.

Sections~\ref{sec:environment}--\ref{sec:action} develop these points in
order. We advocate a research direction; the simulation illustrates the proposed
evaluation and does not test any deployed market.

\section{The Market Belongs in the Learning Problem}
\label{sec:environment}

\subsection{Environment, reward, and credit}
\begin{figure*}[t]
\centering
\includegraphics[width=\textwidth]{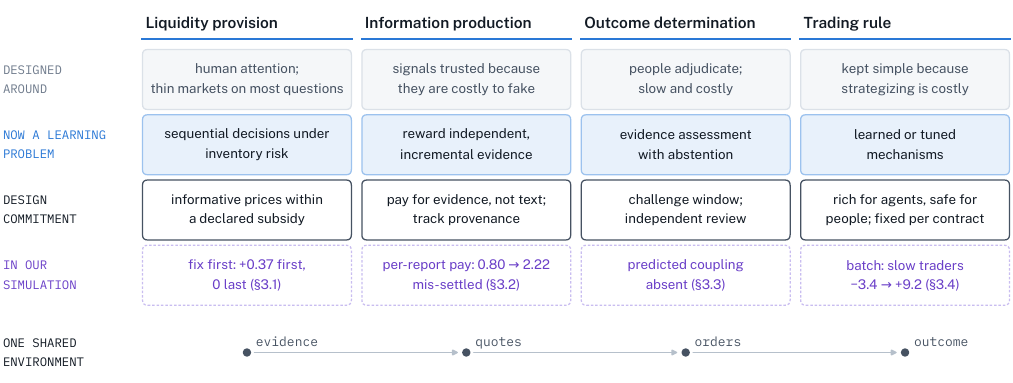}
\caption{\textbf{Each market function, before and after learning agents.} Each
column reads top to bottom: the human limit the function was designed around,
the learning problem it becomes, the design commitment
Section~\ref{sec:functions} argues for, and what our simulation found. The
functions act on one shared environment: evidence informs quotes, quotes shape
orders, and resolved outcomes feed evaluation.}
\label{fig:hypermarket}
\end{figure*}

For an agent that learns in a market, three things that are usually separate
coincide. The environment is non-stationary because other participants learn
too. The reward is the market's payoff rule. The data are prices that other
agents' actions produced. Multi-agent learning treats co-learners as part of
the environment, and two-level formulations learn a policy for the
environment's designer alongside the agents' policies
\citep{zheng2022economist}. Environments that respond to deployed models are
studied as performative prediction \citep{perdomo2020performative} and
strategic classification \citep{hardt2016strategic}; a market whose
participants learn is such an environment, with the added feature that its
rule is also a reward. Financial-agent research rarely adopts this view: the
market is inherited and held fixed, and the score credits the agent with an
outcome it shares with its market, much as people explain behavior by the
person rather than the situation \citep{ross1977intuitive}. 

A market also does more than combine answers that already exist. It creates
incentives to acquire information, determines when participants can act on
it, allocates exposure, and specifies how claims become payments. An ensemble
receives a collection of forecasts; a market can influence which forecasts are
worth producing. Information acquisition must be rewarded, which is why prices
can reveal only part of the available information
\citep{grossman1980impossibility}, and strategic aggregation depends on the
structure of the market \citep{ostrovsky2012information}.

The functional perspective on financial systems separates what must be done
from the institution that happens to do it
\citep{merton1995functional,merton2005design}. We organize the learning
problem around liquidity provision, information production, outcome
determination, and the trading rule. Price discovery results from these
functions and participation, rather than being a fifth component to optimize
separately. The same four functions exist in securities markets under other names. Dealers and designated market
makers provide liquidity; analysts, rating agencies, and data vendors produce information; clearing
houses settle trades, and for credit derivatives industry committees decide whether a credit event
occurred; exchanges set matching rules, which can be redesigned, as frequent batch auctions show
\citep{budish2015hft}. Prediction markets make the argument easiest to state because their claims
resolve cleanly, but the questions carry over: which of these functions learning agents now perform,
what they are rewarded for, and how the people on the other side are protected. Securities markets, with their regulation and systemic risk, are the next setting rather than the first.

An agent's objective and a market's objective are distinct. A dealer could
earn more by widening spreads while making information costly to express; a
forecaster could be accurate but redundant with every other forecaster; an
automated resolver could be fast while excluding evidence it cannot process.
Choosing the rule that governs everyone's orders lies outside a trader's action space; the proposed shift adds such choices to what financial AI builds.

\paragraph{The design problem.} Let $\theta$ collect the designer's choices (quoting policy, evidence
rewards, settlement procedure, clearing rule), restricted to a set $\Theta_C$ that encodes fixed
commitments such as contract terms, exposure limits, update schedules, and dispute rights. Given $\theta$,
people $h \in H$ and agents $a \in A$ choose or learn policies whose joint response $\pi^*(\theta)$ is an
equilibrium or the limit of the learning dynamics. The research problem is
\begin{equation}
\max_{\theta \in \Theta_C} \; J\big(\theta, \pi^*(\theta)\big)
\quad \text{s.t.} \quad U_h\big(\theta, \pi^*(\theta)\big) \ge \underline{U}_h \;\; \forall h \in H,
\end{equation}
where $J$ is the declared market-level objective, for example the proper-score accuracy of prices net of
subsidy and review costs, and each person's outcome must not fall below a reference level
$\underline{U}_h$, such as their outcome in the existing market. Agent-centered research fixes $\theta$
and improves one agent's policy. Treating $\theta$ as a variable is what makes the problem bilevel, and
the inner response $\pi^*(\theta)$ is what makes it a learning problem rather than a static design.

\subsection{What is new, and what is not}
\providecommand{\singlefigwidth}{\columnwidth}
\begin{figure}[!htb]
\centering
\includegraphics[width=\singlefigwidth]{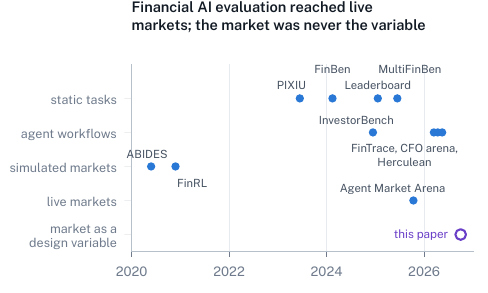}
\caption{\textbf{What financial AI evaluation varies.} Benchmarks by what they
score, dated by first public version: static tasks
\citep{xie2023pixiu,xie2024finben,lin2025openfinllm,peng2025multifinben},
agent workflows \citep{li2024investorbench,han2026cfo,cao2026fintrace,peng2026herculean},
simulated markets \citep{byrd2020abides,liu2021finrl}, and live markets
\citep{qian2025agentstrade}. In each, the market is held fixed.}
\label{fig:timeline}
\end{figure}

Market design already treats institutions as objects to be engineered
\citep{roth2002engineer}. Automated market makers provide rule-based
liquidity \citep{hanson2007lmsr,angeris2020improved}, convex market making
connects that problem to online learning \citep{abernethy2013efficient}, and
learned auctions and two-level policy learning
\citep{duetting2019optimal,zheng2022economist} build on automated mechanism
design \citep{conitzer2002complexity}, so learning a mechanism is an
established idea.  What is new is an agent that strategizes as fully as
a rule rewards and that can perform market functions rather than only bid.
HyperMarket research extends and connects these lines of work for such agents. Financial AI evaluation has followed the same path from models to agents
(Figure~\ref{fig:timeline}): it began with static tasks, moved to tool-using
workflows and simulated firms, and now runs agents in live markets. Each step
widened what the agent does while holding the market fixed. The next step is to make the market's functions the object of design. A recent central-bank review calls the design of rules for markets populated by AI agents an important and largely unstudied question \citep{cartea2026ecb}; auction format and platform rules already change whether algorithms collude \citep{banchio2022auction,johnson2023platform}.

Joint design means that each function is designed and evaluated with the
responses of the others and of the participants in view, as the inner
response $\pi^*(\theta)$ expresses, rather than that one model controls every
function or that human preferences become training parameters. Learning is
also compatible with stable rules, which $\Theta_C$ encodes: a policy can be
trained offline, tested, and held fixed for a contract's lifetime, with
adaptation confined to declared parameters and update intervals. An ordinary exchange with an
LLM trader falls outside our definition, since only the trader learns.

\subsection{What a HyperMarket should be better at}

A HyperMarket might expand the range of claims it can sustain under a stated budget, the
quality of information its prices convey, and the conditions under which
people can contribute and benefit. A claim of progress must identify the capability gained, the costs paid, and whose outcomes
changed.

\section{Building the Functions of a HyperMarket}
\label{sec:functions}
\providecommand{\singlefigwidth}{\columnwidth}
\newcommand{\figliquidity}{\begin{figure}[!htb]
\centering
\includegraphics[width=\singlefigwidth]{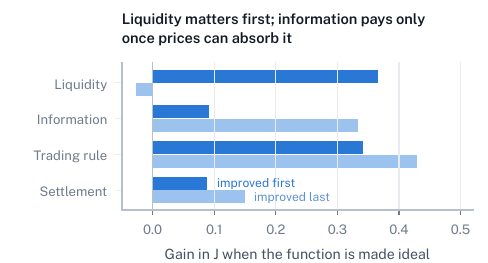}
\caption{\textbf{Liquidity first.} Gain in the market objective $J$ when one function of the fixed design is made ideal, improved first (dark) or last (light); 40 held-out worlds.}
\label{fig:liquidity}
\end{figure}
}
\newcommand{\figinformation}{\begin{figure}[!htb]
\centering
\includegraphics[width=\singlefigwidth]{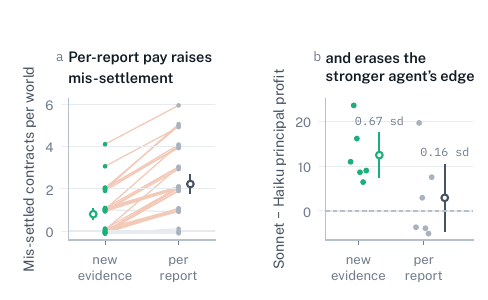}
\caption{\textbf{What the monitor is paid for.} (a) Mis-settled contracts per world, 40 stylized worlds; red lines rise. (b) Profit of principals represented by Sonnet minus Haiku in six worlds with Claude agents reading the news, with the gap in standard deviations of individual profit.}
\label{fig:information}
\end{figure}
}
\newcommand{\figsettlement}{\begin{figure}[!htb]
\centering
\includegraphics[width=\singlefigwidth]{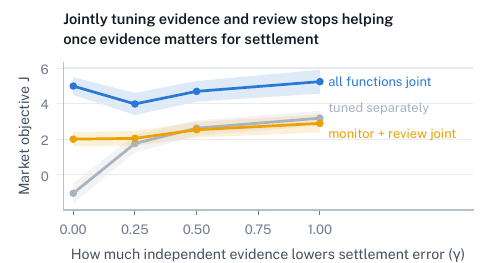}
\caption{\textbf{The coupling we predicted.} $J$ as evidence becomes more useful for settlement; each design re-tuned at each $\gamma$; bands are 95\% intervals.}
\label{fig:settlement}
\end{figure}
}
\newcommand{\figrule}{\begin{figure}[!htb]
\centering
\includegraphics[width=\singlefigwidth]{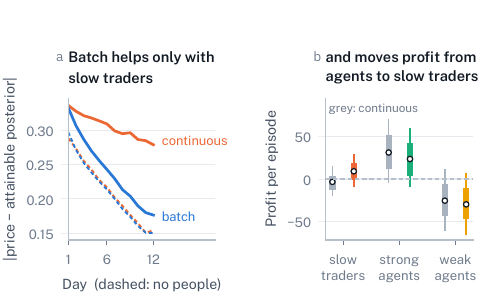}
\caption{\textbf{Batch clearing in a mixed market.} (a) Distance of price from the posterior attainable from all evidence, with (solid) and without (dashed) slow traders. (b) Profit per episode, continuous (grey) and batch (colour); bars span the 10th--90th percentiles, circles are means.}
\label{fig:rule}
\end{figure}
}
\newcommand{\figjoint}{\begin{figure}[!htb]
\centering
\includegraphics[width=\singlefigwidth]{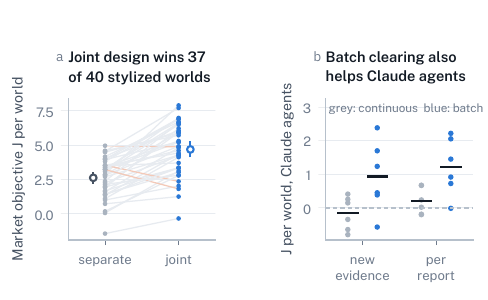}
\caption{\textbf{Joint design under one budget.} (a) $J$ per world, functions tuned separately or jointly; red lines are worlds where joint design loses; circles are means with 95\% intervals. (b) $J$ per world with Claude agents under continuous (grey) and batch (blue) trading.}
\label{fig:joint}
\end{figure}
}

Each function in Figure~\ref{fig:hypermarket} was designed around a limit of
human participants that learning agents relax. Each subsection gives the
evidence on that limit, the learning problem the function becomes, and the
design commitment it requires, and ends with a result from one pre-registered
simulation (Appendix~\ref{app:sim}). In it, a sponsor with a fixed budget
funds monitoring, liquidity, and settlement review for up to 60
dataset-release contracts, traded for 12 days by principals represented by
strong or weak agents and by slow traders. We report 40 held-out worlds,
and six in which Claude Sonnet and Haiku~4.5 read the news.

\subsection{Liquidity provision: make information actionable}
\label{sec:making}
\figliquidity

Prediction markets remain thin on all but prominent questions, because human
liquidity follows human attention \citep{wolfers2004prediction,arrow2008promise},
and information moves a price only if someone can trade on it at acceptable
cost. The logarithmic market scoring rule offers bounded-loss automated
quoting \citep{hanson2007lmsr}, and learned dealers treat quoting as
sequential decision-making under inventory risk \citep{spooner2018market}, and language models can now serve as the policies of hierarchical trading learners \citep{giannouris2026moira}. The
HyperMarket question is how to direct that learning toward informative prices
at a declared subsidy, rather than toward the dealer's profit. A release-date
market could spread a fixed support budget over a contract's life according to
evidence arrival, with a fixed-liquidity market maker as comparator. Quotes
should adapt to declared public signals, not to inferred human vulnerability,
and the learned action set must preserve the scoring rule's loss bound.

\emph{In simulation}, liquidity is the function to fix first: making it ideal
adds 0.37 to the market objective $J$ when done first and nothing when done
last, while better information adds 0.09 first and 0.33 last
(Figure~\ref{fig:liquidity}). A component evaluated alone is misvalued.

\subsection{Information production: pay for evidence, not text}
\label{sec:information}
\figinformation

Language-model forecasters already rival human crowds, and averaging human and machine forecasts was more accurate still \citep{schoenegger2024wisdom}; retrieval improves their reading of financial time series \citep{xiao2025rafts}. Evidence from equity markets suggests that AI adoption can change the informational environment in which trading agents operate: \citet{lopezlira2026chatgpt} show that GPT-4 predicts returns from firm-specific news, but this predictability declines as LLM adoption expands. As AI reduces information-processing frictions, the value of processing certain types of information may decline, changing incentives for information acquisition and trading. This motivates treating market mechanisms and information-production incentives as endogenous to the capabilities and adoption of AI agents. Yet
markets have trusted signals that are costly to fake, which generative text weakens \citep{galdin2025talk}. Trading early on information that will be disclosed anyway earns private profit with little social value, unlike discovering what would otherwise stay hidden \citep{hirshleifer1971private}, and no current market mechanism aligns the two \citep{cartea2026ecb}. The binding constraint can also be what agents
know about their principals: when agents traded books for 201 people,
participants ended at 0.55 on their own rankings against an attainable 0.89,
and 85\% of the shortfall came from imprecise representation of preferences \citep{anthropic2026swap}; in financial advice, what investors chose and what serves their stated risk preferences can diverge \citep{wang2026convfinre}. The learning problem is to reward the incremental
contribution of evidence, accounting for cost, duplication, and strategic
withholding, rather than the volume of plausible text; peer prediction
\citep{miller2005peer} and the Bayesian truth serum \citep{prelec2004bts}
reward reports without ground truth. Agreement among copies of one agent
counts as one source, and shared foundation models make such dependence a
first-order concern \citep{bommasani2022picking,kleinberg2021monoculture}.

\emph{In simulation}, paying the monitor per report floods the record with
re-reports and raises mis-settled contracts from 0.80 to 2.22 per world, in 34
of 40 worlds (Figure~\ref{fig:information}a). With Claude agents, the same
choice decides who gains from a better agent. When the monitor is paid for new
evidence, principals represented by Sonnet earn $12.5 \pm 2.6$ more per
episode than those represented by Haiku in all six worlds, a gap of 0.67
standard deviations of individual profit that few would notice, as Project
Deal found; when it is paid per report, the gap is $2.9 \pm 3.9$
(Figure~\ref{fig:information}b).

\subsection{Outcome determination: connect claims to the world}
\label{sec:determination}
\figsettlement

Settlement has been performed by people, and crowdsourced and decentralized
oracle designs study how to organize it \citep{freeman2017crowdsourced,
caldarelli2020understanding}. Agents can gather and organize evidence while people adjudicate contested interpretations, but checking claims against structured filings and verifying each reasoning step remain open problems for current models \citep{wang2025finauditing,xie2025finchain}. The learning problem is evidence
assessment with abstention under adversarial inputs: interested participants
may publish misleading evidence or place instructions in documents, so
resolvers must be tested against fabricated and conflicting sources \citep{liu2024fmdllama,jiang2026glisters}, with
trading and adjudication separated. The dataset contract should specify
before trading what counts as a release and which sources are admissible, and
a challenge window determines when a proposed outcome becomes final. Some
contracts create incentives to cause the event itself
\citep{chakraborty2016trading} and should stay unlisted.

\emph{In simulation}, we expected evidence to substitute for review, so that
tuning monitoring and review together would pay more as evidence mattered
more for settlement. It did not: that pair beats separate tuning only when
evidence is irrelevant to settlement (Figure~\ref{fig:settlement}), because a
reviewer tuned for the worst case still benefits from evidence when it
arrives.

\subsection{Trading rules: rich enough for agents, safe for people}
\label{sec:mechanism}
\figrule

Trading rules have been kept simple largely because strategizing is costly
for people: eBay moved from auctions to posted prices
\citep{einav2018auctions,fisman2016inner}, and school districts adopted
strategy-proof assignment partly to protect sincere families from
sophisticated ones \citep{pathak2008leveling}. Learning agents make richer
mechanisms usable, and in a mixed market they are the sophisticated players.
Items sold by a stronger model fetched \$3.64 more than when sold by a weaker
one, yet participants rated the fairness of the two models' deals almost
identically \citep{anthropic2026deal}. Time is part of the problem:
continuous trading rewards speed \citep{budish2015hft}, and on decentralized
exchanges ordering rules let bots extract value from other users
\citep{daian2020flash}. Investors already vote with their orders: on Euronext Amsterdam, the share of non-proprietary volume in large-cap stocks traded on the continuous order book fell from 65\% to under 45\% between 2020 and 2025 as trading moved to the closing auction \citep{cartea2026ecb}. Periodic batch clearing is one response; Feeding
America's daily sealed-bid rounds let food banks without staff to watch
auctions compete \citep{prendergast2017foodbanks}. Participants must be able
to anticipate and understand the rule, so updates belong between contracts,
on a public schedule.

\emph{In simulation}, uniform-price batch clearing brings prices closer to the
attainable posterior and turns slow traders' average result from $-3.4$ to
$+9.2$, but only when slow traders are present; with agents alone the rules
coincide (Figure~\ref{fig:rule}). With Claude agents, whose beliefs are more
moderate, batch clearing still raises $J$ but no longer protects slow traders
($-0.83 \pm 0.63$).

\figjoint
\subsection{Joint design pays through interactions between functions}
\label{sec:joint}

Tuning all four functions jointly for the market objective, rather than each
for its own, raises $J$ from 2.62 to 4.69 under the same budget (paired
difference $2.07 \pm 0.22$; 37 of 40 worlds; Figure~\ref{fig:joint}a), and
batch clearing raises $J$ with Claude agents as well ($+1.09 \pm 0.27$, six of
six worlds; Figure~\ref{fig:joint}b). The gain does not come from the channel
we predicted (Section~\ref{sec:determination}). It comes from a rule whose best
setting depends on who is in the market, among functions whose values depend
on one another. These results show that such interactions exist and can be
measured, not their size in any real market.

Joint design can also create dangerous dependencies: if one
operator profits from trading and controls the outcome label, optimizing its
total reward can undermine truthful settlement. Independent adjudication and
separated access remain constraints even when a unified model would be
cheaper; the objective is coordination among accountable functions rather
than consolidation of authority in one learner.

\section{Human--AI Collective Intelligence Is a Design Objective}
\label{sec:aggregation}

\subsection{Participation is not yet complementarity}

People can contribute local knowledge, different access to information, and
judgments about what makes a claim meaningful. Agents can contribute
continuous monitoring and cheap replication of some analyses.  Since human--AI combinations have on average underperformed the better of
their parts (Figure~\ref{fig:evidence}c), a demonstration of complementarity
must compare the combination with strong alternatives
\citep{vaccaro2024combinations,hemmer2021human}. Assistance also differs from market-level complementarity: a tool that improves each person's forecast \citep{schoenegger2024ai} may leave a population's information unchanged if everyone receives the same advice.

For prediction tasks, a first operational target is the quality of market
probabilities measured against independently resolved outcomes with a
declared proper scoring rule \citep{gneiting2007strictly}. Fairness, accessibility, and benefit to participants require separate measures: language-model credit scoring can help many applicants while biasing outcomes for some \citep{feng2023credit}, and financial language models lag in Spanish and Greek relative to English \citep{zhang2024dolares,peng2025plutus}, which bears on who can take part in a market run through them.

\subsection{What would support or weaken the position}
\providecommand{\singlefigwidth}{\columnwidth}
\begin{figure}[!htb]
\centering
\includegraphics[width=\singlefigwidth]{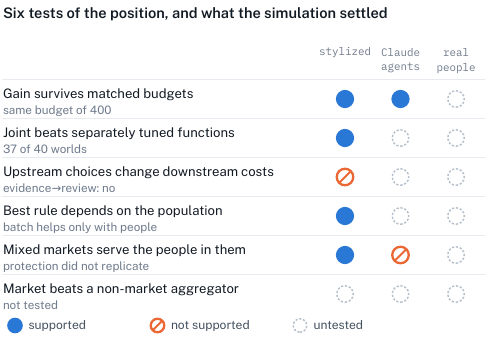}
\caption{\textbf{What would support or weaken the position.} Each row is a
test; each column a population. Our simulation settles some tests for
stylized and Claude agents; none has been run with real people.}
\label{fig:criteria}
\end{figure}

The research should separate improvements in agents from improvements in
institutions: hold agent capability and evidence access comparable while
varying the mechanism, vary the population across human-only, AI-only, and
mixed participation, and include a competitive non-market aggregator. All
comparisons should report information, compute, subsidy, and human-time
budgets, since a market that buys more evidence has found more evidence rather
than aggregated it better. Figure~\ref{fig:criteria} lists six tests and what
our simulation settled. The position is weakened wherever gains vanish under
matched budgets, the best design does not depend on the population, an
ensemble matches the market at lower cost, or mixed participation worsens
outcomes for people. Benefits to people must be measured on people, and a
single success would establish possibility rather than broad applicability.

\section{Alternative Views}
\label{sec:alternative}

\paragraph{Improve and certify agents; leave institutions stable.}
Better forecasting, reliable tool use, and certification can address many
failures without changing market infrastructure, and evidence of tacit
collusion argues for certifying agents before they make market decisions
\citep{collusion2026certification}. The agendas are complementary: the
first live agent marketplaces pair tests of agents with market rules for
registration, rate limits, and failed deals \citep{anthropic2026swap}. But an
agent's behavior depends on the environment and population in which it acts,
so better agents make better markets only under suitable rules. Adverse
selection unravels a market precisely when buyers reason correctly about it
\citep{akerlof1970market,fisman2016inner}, so fully rational agents could
thin some markets rather than improve them; in our simulation, the same
agents gained or lost depending on the rule (Section~\ref{sec:joint}). Certification also
leaves open design questions such as how to reward information production or
support an illiquid claim. Where certified agents under existing rules
suffice for an objective, the existing rules should stay.

\paragraph{This is existing mechanism design with a new name.}
Learned auctions, adaptive market making, the AI Economist, and artificial
prediction markets already cover parts of the proposal
\citep{duetting2019optimal,spooner2018market,zheng2022economist,
chakravorti2022artificial}. The position concerns what financial AI should build and
connect: agent-operated market functions studied together with the mixed
populations they serve. Its value depends on finding interactions that
isolated component studies miss; if the functions can be designed
independently without consequential loss, the case for an integrated agenda
is weaker. 

\paragraph{Learning makes mechanisms less trustworthy.}
Fixed rules can be understood, audited, and anticipated, while a learned
policy may be opaque, unstable, or exploitable through its updates. The
proposal is compatible with offline learning behind fixed action constraints,
human dispute procedures, and rules simple enough to follow; where a benefit
fails to survive these restrictions, the fixed alternative should win.

\paragraph{Use ensembles or deliberation instead of markets.}
Forecasting ensembles already rival human crowds without budgets, strategy,
or transaction costs. The case for a market is strongest when information is
costly, distributed, and strategically supplied, and weakest when judgments
can be collected cheaply from cooperative contributors.

\paragraph{Sufficiently capable agents make human participation unnecessary.}

People remain affected parties with purposes, rights, and information they may
choose to withhold. Deciding whose interests count
requires more than greater capability.

\section{Call to Action}
\label{sec:action}

\paragraph{Researchers: make the mechanism a variable.}
A practical first project combines one learned function with fixed
contracts and a strong baseline. State the intended market benefit before
choosing the agent's reward, and publish cases where fixed rules or non-market methods
remain preferable. Later projects should study interactions between functions.

\paragraph{Benchmark and simulator builders: expose the market.} Existing financial suites and leaderboards \citep{xie2024finben,lin2025openfinllm} and agent simulators with budgets and delayed consequences \citep{han2026cfo,qian2025agentstrade} are natural hosts for a market track.
Provide separable interfaces for information acquisition, quoting, clearing,
and settlement, including costs and disputes. Treat the population as a reported variable: results should state the
mechanism, population, and shared base models, and be repeated across
populations.

\paragraph{Venues and reviewers: recognize institutional progress.}
Invite results that improve a market even when its traders are unchanged, and ask of trading-agent results in which market
they were obtained.  

\paragraph{Operators: provide bounded, accountable test settings.}
Early studies should use simulation and consenting research participants
before any claim about live deployment. Pilots should specify contracts, capital limits, and appeal procedures, and reports should separate better
forecasting from transfers between participant groups. 

\section{Conclusion}

Machine learning has made the market's traders into learners. It can now make
the rest of the market into learners as well: the functions that provide
liquidity, produce information, settle claims, and set the rules. Holding
those functions fixed while improving traders decides, without discussion, the
environment and the reward for a growing class of agents. HyperMarkets make the alternative
concrete: market functions performed or adapted by learning agents, designed
with human and AI participants toward objectives stated at the level of the
market. The opportunity is to sustain claims that human attention never
reached and to use information markets have failed to elicit. The
challenge is to show that the resulting coordination is worth its costs and
serves the people in it. Human--AI collective intelligence is the objective to
be established, not a benefit that follows from more capable agents.

\section*{Impact Statement}
The proposed research concerns incentives and institutions that could affect
people's resources, information, and opportunities. Improved forecast accuracy
does not justify extracting value from less informed participants, collecting
unnecessary personal data, or creating harmful incentives around an event.
Studies involving people require appropriate ethics review, informed consent,
declared exposure limits, and mechanisms for disputes and withdrawal. Contracts
whose incentives encourage harm should not be listed. This paper reports no
new human-subject study or market deployment. Its proposed safeguards are
design commitments, not evidence that a future implementation is safe.

\section*{AI Use Statement}
OpenAI Codex and Anthropic Claude assisted with reviewing related literature,
revising the manuscript, preparing figures and document-processing scripts, and
implementing the simulation. In the simulation (Appendix~\ref{app:sim}), Claude
Sonnet and Haiku~4.5 act as experimental participants that read generated news
and report probabilities; these outputs are simulation data, not evidence about
people or live markets.

\bibliographystyle{plainnat}
\bibliography{refs}

\begin{thebibliography}{76}
\providecommand{\natexlab}[1]{#1}
\providecommand{\url}[1]{\texttt{#1}}
\expandafter\ifx\csname urlstyle\endcsname\relax
  \providecommand{\doi}[1]{doi: #1}\else
  \providecommand{\doi}{doi: \begingroup \urlstyle{rm}\Url}\fi

\bibitem[Abernethy et~al.(2013)Abernethy, Chen, and
  Vaughan]{abernethy2013efficient}
Jacob Abernethy, Yiling Chen, and Jennifer~Wortman Vaughan.
\newblock {Efficient Market Making via Convex Optimization, and a Connection to
  Online Learning}.
\newblock \emph{ACM Transactions on Economics and Computation}, 1\penalty0
  (2):\penalty0 1--39, 2013.
\newblock \doi{https://doi.org/10.1145/2465769.2465777}.

\bibitem[Akerlof(1970)]{akerlof1970market}
George~A. Akerlof.
\newblock The market for ``lemons'': Quality uncertainty and the market
  mechanism.
\newblock \emph{The Quarterly Journal of Economics}, 84\penalty0 (3):\penalty0
  488--500, 1970.
\newblock \doi{10.2307/1879431}.

\bibitem[Angeris and Chitra(2020)]{angeris2020improved}
Guillermo Angeris and Tarun Chitra.
\newblock {Improved Price Oracles: Constant Function Market Makers}.
\newblock SSRN Electronic Journal, 2020.

\bibitem[Arrow et~al.(2008)Arrow, Forsythe, Gorham, Hahn, Hanson, Ledyard,
  Levmore, Litan, Milgrom, and Nelson]{arrow2008promise}
Kenneth~J. Arrow, Robert Forsythe, Michael Gorham, Robert Hahn, Robin Hanson,
  John~O. Ledyard, Saul Levmore, Robert Litan, Paul Milgrom, and Forrest~D.
  Nelson.
\newblock {The Promise of Prediction Markets}.
\newblock \emph{Science}, 2008.
\newblock \doi{10.1126/science.1157679}.

\bibitem[Banchio and Skrzypacz(2022)]{banchio2022auction}
Martino Banchio and Andrzej Skrzypacz.
\newblock Artificial intelligence and auction design.
\newblock In \emph{Proceedings of the 23rd ACM Conference on Economics and
  Computation}, pages 30--31, 2022.
\newblock \doi{10.1145/3490486.3538244}.

\bibitem[Bommasani et~al.(2022)Bommasani, Creel, Kumar, Jurafsky, and
  Liang]{bommasani2022picking}
Rishi Bommasani, Kathleen~A. Creel, Ananya Kumar, Dan Jurafsky, and Percy
  Liang.
\newblock {Picking on the Same Person: Does Algorithmic Monoculture Lead to
  Outcome Homogenization?}
\newblock In \emph{Advances in Neural Information Processing Systems 35}, 2022.
\newblock \doi{10.52202/068431-0265}.

\bibitem[Budish et~al.(2015)Budish, Cramton, and Shim]{budish2015hft}
Eric Budish, Peter Cramton, and John Shim.
\newblock The high-frequency trading arms race: Frequent batch auctions as a
  market design response.
\newblock \emph{The Quarterly Journal of Economics}, 130\penalty0 (4):\penalty0
  1547--1621, 2015.
\newblock \doi{10.1093/qje/qjv027}.

\bibitem[Byrd et~al.(2020)Byrd, Hybinette, and Balch]{byrd2020abides}
David Byrd, Maria Hybinette, and Tucker~Hybinette Balch.
\newblock {ABIDES}: Towards high-fidelity multi-agent market simulation.
\newblock In \emph{Proceedings of the 2020 ACM SIGSIM Conference on Principles
  of Advanced Discrete Simulation}, pages 11--22, 2020.
\newblock \doi{10.1145/3384441.3395986}.

\bibitem[Caldarelli(2020)]{caldarelli2020understanding}
Giulio Caldarelli.
\newblock {Understanding the Blockchain Oracle Problem: A Call for Action}.
\newblock \emph{Information}, 11\penalty0 (11):\penalty0 509, 2020.
\newblock \doi{https://doi.org/10.3390/info11110509}.

\bibitem[Calvano et~al.(2020)Calvano, Calzolari, Denicol{\`o}, and
  Pastorello]{calvano2020collusion}
Emilio Calvano, Giacomo Calzolari, Vincenzo Denicol{\`o}, and Sergio
  Pastorello.
\newblock Artificial intelligence, algorithmic pricing, and collusion.
\newblock \emph{American Economic Review}, 110\penalty0 (10):\penalty0
  3267--3297, 2020.
\newblock \doi{10.1257/aer.20190623}.

\bibitem[Cao et~al.(2026)Cao, Li, Liu, Cao, Xu, Qian, Peng, Tang, Yao, Huang,
  et~al.]{cao2026fintrace}
Yupeng Cao, Haohang Li, Weijin Liu, Wenbo Cao, Anke Xu, Lingfei Qian, Xueqing
  Peng, Minxue Tang, Zhiyuan Yao, Jimin Huang, et~al.
\newblock {FinTrace}: Holistic trajectory-level evaluation of {LLM} tool
  calling for long-horizon financial tasks, 2026.

\bibitem[Cartea et~al.(2026{\natexlab{a}})Cartea, Chang, Penalva, and
  Waldon]{cartea2026folk}
{\'A}lvaro Cartea, Patrick Chang, Jos{\'e} Penalva, and Harrison Waldon.
\newblock Algorithmic collusion and a folk theorem from learning with bounded
  rationality.
\newblock \emph{Games and Economic Behavior}, 157:\penalty0 1--21,
  2026{\natexlab{a}}.
\newblock \doi{10.1016/j.geb.2025.11.012}.

\bibitem[Cartea et~al.(2026{\natexlab{b}})Cartea, Colliard, Foucault, Hoffmann,
  Graumans, and Sigaux]{cartea2026ecb}
{\'A}lvaro Cartea, Jean-Edouard Colliard, Thierry Foucault, Peter Hoffmann, Rob
  Graumans, and Jean-David Sigaux.
\newblock Artificial intelligence and financial markets.
\newblock ECB Discussion Paper Series~32, European Central Bank,
  2026{\natexlab{b}}.

\bibitem[Chakraborty and Das(2016)]{chakraborty2016trading}
Mithun Chakraborty and Sanmay Das.
\newblock {Trading on a rigged game: outcome manipulation in prediction
  markets}.
\newblock In \emph{Proceedings of the Twenty-Fifth International Joint
  Conference on Artificial Intelligence}, pages 158--164, 2016.
\newblock URL \url{https://www.ijcai.org/Proceedings/16/Papers/030.pdf}.

\bibitem[Chakravorti et~al.(2022)Chakravorti, Singh, Rajtmajer, McLaughlin,
  Fraleigh, Griffin, Kwasnica, Pennock, and Giles]{chakravorti2022artificial}
Tatiana Chakravorti, Vaibhav Singh, Sarah Rajtmajer, Michael McLaughlin, Robert
  Fraleigh, Christopher Griffin, Anthony Kwasnica, David Pennock, and C.~Lee
  Giles.
\newblock Artificial prediction markets present a novel opportunity for
  human-{AI} collaboration.
\newblock arXiv preprint arXiv:2211.16590, 2022.
\newblock URL \url{https://arxiv.org/abs/2211.16590}.

\bibitem[Conitzer and Sandholm(2002)]{conitzer2002complexity}
Vincent Conitzer and Tuomas Sandholm.
\newblock Complexity of mechanism design.
\newblock In \emph{Proceedings of the 18th Conference on Uncertainty in
  Artificial Intelligence}, 2002.
\newblock \doi{10.48550/arXiv.cs/0205075}.

\bibitem[Daian et~al.(2020)Daian, Goldfeder, Kell, Li, Zhao, Bentov,
  Breidenbach, and Juels]{daian2020flash}
Philip Daian, Steven Goldfeder, Tyler Kell, Yunqi Li, Xueyuan Zhao, Iddo
  Bentov, Lorenz Breidenbach, and Ari Juels.
\newblock Flash boys 2.0: Frontrunning in decentralized exchanges, miner
  extractable value, and consensus instability.
\newblock In \emph{2020 IEEE Symposium on Security and Privacy}, pages
  910--927, 2020.
\newblock \doi{10.1109/SP40000.2020.00040}.

\bibitem[Duetting et~al.(2019)Duetting, Feng, Narasimhan, Parkes, and
  Ravindranath]{duetting2019optimal}
Paul Duetting, Zhe Feng, Harikrishna Narasimhan, David Parkes, and Sai~Srivatsa
  Ravindranath.
\newblock Optimal auctions through deep learning.
\newblock In \emph{Proceedings of the 36th International Conference on Machine
  Learning}, volume~97 of \emph{Proceedings of Machine Learning Research},
  pages 1706--1715. PMLR, 2019.
\newblock URL \url{https://proceedings.mlr.press/v97/duetting19a.html}.

\bibitem[Einav et~al.(2018)Einav, Farronato, Levin, and
  Sundaresan]{einav2018auctions}
Liran Einav, Chiara Farronato, Jonathan Levin, and Neel Sundaresan.
\newblock Auctions versus posted prices in online markets.
\newblock \emph{Journal of Political Economy}, 126\penalty0 (1):\penalty0
  178--215, 2018.
\newblock \doi{10.1086/695529}.

\bibitem[Feng et~al.(2023)Feng, Dai, Huang, Zhang, Xie, Han, Chen, Lopez-Lira,
  and Wang]{feng2023credit}
Duanyu Feng, Yongfu Dai, Jimin Huang, Yifang Zhang, Qianqian Xie, Weiguang Han,
  Zhengyu Chen, Alejandro Lopez-Lira, and Hao Wang.
\newblock Empowering many, biasing a few: Generalist credit scoring through
  large language models, 2023.

\bibitem[Fisman and Sullivan(2016)]{fisman2016inner}
Ray Fisman and Tim Sullivan.
\newblock \emph{The Inner Lives of Markets: How People Shape Them---And They
  Shape Us}.
\newblock PublicAffairs, New York, 2016.
\newblock ISBN 9781610394925.

\bibitem[Freeman et~al.(2017)Freeman, Lahaie, and
  Pennock]{freeman2017crowdsourced}
Rupert Freeman, Sébastien Lahaie, and David~M. Pennock.
\newblock {Crowdsourced Outcome Determination in Prediction Markets}.
\newblock In \emph{Proceedings of the AAAI Conference on Artificial
  Intelligence}, volume~31, 2017.
\newblock \doi{https://doi.org/10.1609/aaai.v31i1.10594}.

\bibitem[Galdin and Silbert(2025)]{galdin2025talk}
Ana{\"i}s Galdin and Jesse Silbert.
\newblock Making talk cheap: Generative {AI} and labor market signaling.
\newblock SSRN working paper, 2025.

\bibitem[Giannouris et~al.(2026)Giannouris, Jiang, Qian, Wang, Peng, Huang,
  Xiong, and Ananiadou]{giannouris2026moira}
Polydoros Giannouris, Yuechen Jiang, Lingfei Qian, Yuyan Wang, Xueqing Peng,
  Jimin Huang, Guojun Xiong, and Sophia Ananiadou.
\newblock Moira: Language-driven hierarchical reinforcement learning for pair
  trading, 2026.

\bibitem[Gneiting and Raftery(2007)]{gneiting2007strictly}
Tilmann Gneiting and Adrian~E. Raftery.
\newblock {Strictly Proper Scoring Rules, Prediction, and Estimation}.
\newblock \emph{Journal of the American Statistical Association}, 102\penalty0
  (477):\penalty0 359--378, 2007.
\newblock \doi{https://doi.org/10.1198/016214506000001437}.

\bibitem[Grossman and Stiglitz(1980)]{grossman1980impossibility}
Sanford~J. Grossman and Joseph~E. Stiglitz.
\newblock On the impossibility of informationally efficient markets.
\newblock \emph{American Economic Review}, 70\penalty0 (3):\penalty0 393--408,
  1980.

\bibitem[Halawi et~al.(2024)Halawi, Zhang, Yueh-Han, and
  Steinhardt]{halawi2024forecasting}
Danny Halawi, Fred Zhang, Chen Yueh-Han, and Jacob Steinhardt.
\newblock {Approaching Human-Level Forecasting with Language Models}.
\newblock In \emph{Advances in Neural Information Processing Systems 37}, 2024.
\newblock \doi{10.52202/079017-1598}.

\bibitem[Han et~al.(2026)Han, Wang, Qian, Li, Cao, He, Peng, Shen, Xu, Chen,
  et~al.]{han2026cfo}
Yi~Han, Yan Wang, Lingfei Qian, Haohang Li, Yupeng Cao, Yueru He, Xueqing Peng,
  Nanhan Shen, Yitao Xu, Yankai Chen, et~al.
\newblock Can {LLM} agents be {CFOs}? benchmarking long-horizon resource
  allocation in an uncertain enterprise environment, 2026.

\bibitem[Hanson(2007)]{hanson2007lmsr}
Robin Hanson.
\newblock Logarithmic market scoring rules for modular combinatorial
  information aggregation.
\newblock \emph{The Journal of Prediction Markets}, 1\penalty0 (1):\penalty0
  3--15, 2007.
\newblock \doi{10.5750/jpm.v1i1.417}.

\bibitem[Hardt et~al.(2016)Hardt, Megiddo, Papadimitriou, and
  Wootters]{hardt2016strategic}
Moritz Hardt, Nimrod Megiddo, Christos Papadimitriou, and Mary Wootters.
\newblock Strategic classification.
\newblock In \emph{Proceedings of the 2016 ACM Conference on Innovations in
  Theoretical Computer Science}, pages 111--122, 2016.
\newblock \doi{10.1145/2840728.2840730}.

\bibitem[Hemmer et~al.(2021)Hemmer, Schemmer, Vössing, and
  Kühl]{hemmer2021human}
Patrick Hemmer, Max Schemmer, Michael Vössing, and Niklas Kühl.
\newblock {Human-AI Complementarity in Hybrid Intelligence Systems: A
  Structured Literature Review}.
\newblock In \emph{PACIS 2021 Proceedings}, page~78, 2021.
\newblock URL \url{https://aisel.aisnet.org/pacis2021/78/}.

\bibitem[Hirshleifer(1971)]{hirshleifer1971private}
Jack Hirshleifer.
\newblock The private and social value of information and the reward to
  inventive activity.
\newblock \emph{The American Economic Review}, 61\penalty0 (4):\penalty0
  561--574, 1971.

\bibitem[Hitzig et~al.(2026)Hitzig, Carr, Cotter, Troy, Turman, Massenkoff, and
  McCrory]{anthropic2026swap}
Zo{\"e} Hitzig, Sylvie Carr, Tess Cotter, Kevin Troy, Kyle Turman, Maxim
  Massenkoff, and Peter McCrory.
\newblock Project swap: What happens when agents trade for us?
\newblock Anthropic research report, 2026.
\newblock URL \url{https://www.anthropic.com/research/project-swap}.

\bibitem[Jiang et~al.(2026)Jiang, Liu, Cao, He, Xu, Xu, Deng, Tiwari, Chen,
  Lopez-Lira, et~al.]{jiang2026glisters}
Yuechen Jiang, Zhiwei Liu, Yupeng Cao, Yueru He, Ziyang Xu, Chen Xu, Zhiyang
  Deng, Prayag Tiwari, Xi~Chen, Alejandro Lopez-Lira, et~al.
\newblock All that glisters is not gold: A benchmark for reference-free
  counterfactual financial misinformation detection, 2026.

\bibitem[Johnson et~al.(2023)Johnson, Rhodes, and
  Wildenbeest]{johnson2023platform}
Justin~P. Johnson, Andrew Rhodes, and Matthijs Wildenbeest.
\newblock Platform design when sellers use pricing algorithms.
\newblock \emph{Econometrica}, 91\penalty0 (5):\penalty0 1841--1879, 2023.
\newblock \doi{10.3982/ECTA19978}.

\bibitem[Kleinberg and Raghavan(2021)]{kleinberg2021monoculture}
Jon Kleinberg and Manish Raghavan.
\newblock Algorithmic monoculture and social welfare.
\newblock \emph{Proceedings of the National Academy of Sciences}, 118\penalty0
  (22), 2021.
\newblock \doi{10.1073/pnas.2018340118}.

\bibitem[Li et~al.(2024)Li, Cao, Yu, Javaji, Deng, He, Jiang, Zhu,
  Subbalakshmi, Xiong, et~al.]{li2024investorbench}
Haohang Li, Yupeng Cao, Yangyang Yu, Shashidhar~Reddy Javaji, Zhiyang Deng,
  Yueru He, Yuechen Jiang, Zining Zhu, Koduvayur Subbalakshmi, Guojun Xiong,
  et~al.
\newblock {INVESTORBENCH}: A benchmark for financial decision-making tasks with
  {LLM-based} agent, 2024.

\bibitem[Lin et~al.(2025)Lin, Tian, Wang, Zhao, Huang, Xie, Borella, White,
  Wang, Xiao, et~al.]{lin2025openfinllm}
Shengyuan~Colin Lin, Felix Tian, Keyi Wang, Xingjian Zhao, Jimin Huang,
  Qianqian Xie, Luca Borella, Matt White, Christina~Dan Wang, Kairong Xiao,
  et~al.
\newblock Open {FinLLM} leaderboard: Towards financial {AI} readiness, 2025.

\bibitem[Liu et~al.(2020)Liu, Yang, Chen, Zhang, Yang, Xiao, and
  Wang]{liu2021finrl}
Xiao-Yang Liu, Hongyang Yang, Qian Chen, Runjia Zhang, Liuqing Yang, Bowen
  Xiao, and Christina~Dan Wang.
\newblock {FinRL: A Deep Reinforcement Learning Library for Automated Stock
  Trading in Quantitative Finance}.
\newblock arXiv preprint arXiv:2011.09607, 2020.
\newblock URL \url{https://arxiv.org/abs/2011.09607}.

\bibitem[Liu et~al.(2024)Liu, Zhang, Yang, Xie, Huang, and
  Ananiadou]{liu2024fmdllama}
Zhiwei Liu, Xin Zhang, Kailai Yang, Qianqian Xie, Jimin Huang, and Sophia
  Ananiadou.
\newblock {FMDLlama}: Financial misinformation detection based on large
  language models, 2024.

\bibitem[Lopez-Lira and Tang(2026)]{lopezlira2026chatgpt}
Alejandro Lopez-Lira and Yuehua Tang.
\newblock Can {ChatGPT} forecast stock price movements? return predictability
  and large language models.
\newblock \emph{Journal of Financial Economics}, 184:\penalty0 104335, 2026.
\newblock \doi{10.1016/j.jfineco.2026.104335}.

\bibitem[Merton and Bodie(2004)]{merton2005design}
Robert Merton and Zvi Bodie.
\newblock {The Design of Financial Systems: Towards a Synthesis of Function and
  Structure}.
\newblock Technical Report 10620, National Bureau of Economic Research, 2004.

\bibitem[Merton(1995)]{merton1995functional}
Robert~C. Merton.
\newblock {A Functional Perspective of Financial Intermediation}.
\newblock \emph{Financial Management}, 24\penalty0 (2):\penalty0 23, 1995.
\newblock \doi{https://doi.org/10.2307/3665532}.

\bibitem[Miller et~al.(2005)Miller, Resnick, and Zeckhauser]{miller2005peer}
Nolan Miller, Paul Resnick, and Richard Zeckhauser.
\newblock Eliciting informative feedback: The peer-prediction method.
\newblock \emph{Management Science}, 51\penalty0 (9):\penalty0 1359--1373,
  2005.
\newblock \doi{10.1287/mnsc.1050.0379}.

\bibitem[Ostrovsky(2012)]{ostrovsky2012information}
Michael Ostrovsky.
\newblock {Information Aggregation in Dynamic Markets With Strategic Traders}.
\newblock \emph{Econometrica}, 80\penalty0 (6):\penalty0 2595--2647, 2012.
\newblock \doi{https://doi.org/10.3982/ecta8479}.

\bibitem[Pathak and S{\"o}nmez(2008)]{pathak2008leveling}
Parag~A. Pathak and Tayfun S{\"o}nmez.
\newblock Leveling the playing field: Sincere and sophisticated players in the
  {Boston} mechanism.
\newblock \emph{American Economic Review}, 98\penalty0 (4):\penalty0
  1636--1652, 2008.
\newblock \doi{10.1257/aer.98.4.1636}.

\bibitem[Peng et~al.(2025{\natexlab{a}})Peng, Papadopoulos, Soufleri,
  Giannouris, Xiang, Wang, Qian, Huang, Xie, and Ananiadou]{peng2025plutus}
Xueqing Peng, Triantafillos Papadopoulos, Efstathia Soufleri, Polydoros
  Giannouris, Ruoyu Xiang, Yan Wang, Lingfei Qian, Jimin Huang, Qianqian Xie,
  and Sophia Ananiadou.
\newblock Plutus: Benchmarking large language models in low-resource {Greek}
  finance, 2025{\natexlab{a}}.

\bibitem[Peng et~al.(2025{\natexlab{b}})Peng, Qian, Wang, Xiang, He, Ren,
  Jiang, Zhang, Guo, Zhao, et~al.]{peng2025multifinben}
Xueqing Peng, Lingfei Qian, Yan Wang, Ruoyu Xiang, Yueru He, Yang Ren, Mingyang
  Jiang, Vincent~Jim Zhang, Yuqing Guo, Jeff Zhao, et~al.
\newblock {MultiFinBen}: Benchmarking large language models for multilingual
  and multimodal financial application, 2025{\natexlab{b}}.

\bibitem[Peng et~al.(2026)Peng, Xie, Cao, Li, Qian, Wang, Zhang, He, Ai, Ma,
  et~al.]{peng2026herculean}
Xueqing Peng, Zhuohan Xie, Yupeng Cao, Haohang Li, Lingfei Qian, Yan Wang,
  Vincent~Jim Zhang, Huan He, Xuguang Ai, Linhai Ma, et~al.
\newblock Herculean: An agentic benchmark for financial intelligence, 2026.

\bibitem[Perdomo et~al.(2020)Perdomo, Zrnic, Mendler-D{\"u}nner, and
  Hardt]{perdomo2020performative}
Juan~C. Perdomo, Tijana Zrnic, Celestine Mendler-D{\"u}nner, and Moritz Hardt.
\newblock Performative prediction.
\newblock In \emph{Proceedings of the 37th International Conference on Machine
  Learning}, 2020.
\newblock \doi{10.48550/arXiv.2002.06673}.

\bibitem[Prelec(2004)]{prelec2004bts}
Dra{\v{z}}en Prelec.
\newblock A {Bayesian} truth serum for subjective data.
\newblock \emph{Science}, 306\penalty0 (5695):\penalty0 462--466, 2004.
\newblock \doi{10.1126/science.1102081}.

\bibitem[Prendergast(2017)]{prendergast2017foodbanks}
Canice Prendergast.
\newblock How food banks use markets to feed the poor.
\newblock \emph{Journal of Economic Perspectives}, 31\penalty0 (4):\penalty0
  145--162, 2017.
\newblock \doi{10.1257/jep.31.4.145}.

\bibitem[Qian et~al.(2025)Qian, Peng, Wang, Zhang, He, Smith, Han, He, Li, Cao,
  et~al.]{qian2025agentstrade}
Lingfei Qian, Xueqing Peng, Yan Wang, Vincent~Jim Zhang, Huan He, Hanley Smith,
  Yi~Han, Yueru He, Haohang Li, Yupeng Cao, et~al.
\newblock When agents trade: Live multi-market trading benchmark for {LLM}
  agents, 2025.

\bibitem[Riemer et~al.(2026)Riemer, Tosato, Memarian, Touzel, Berseth, Rish,
  and Dumas]{collusion2026certification}
Matthew Riemer, Tommaso Tosato, Amin Memarian, Maximilian~Puelma Touzel, Glen
  Berseth, Irina Rish, and Guillaume Dumas.
\newblock {Position: Collusion Risks Among AI Reasoning Agents Justify
  Certification Requirements for Making Market Decisions}, 2026.

\bibitem[Ross(1977)]{ross1977intuitive}
Lee Ross.
\newblock The intuitive psychologist and his shortcomings: Distortions in the
  attribution process.
\newblock In \emph{Advances in Experimental Social Psychology}, volume~10,
  pages 173--220. Academic Press, 1977.
\newblock \doi{10.1016/S0065-2601(08)60357-3}.

\bibitem[Roth(2002)]{roth2002engineer}
Alvin~E. Roth.
\newblock {The Economist as Engineer: Game Theory, Experimentation, and
  Computation as Tools for Design Economics}.
\newblock \emph{Econometrica}, 2002.
\newblock \doi{10.1111/1468-0262.00335}.

\bibitem[Schoenegger et~al.(2024)Schoenegger, Tuminauskaite, Park, Bastos, and
  Tetlock]{schoenegger2024wisdom}
Philipp Schoenegger, Indre Tuminauskaite, Peter~S. Park, Rafael Valdece~Sousa
  Bastos, and Philip~E. Tetlock.
\newblock {Wisdom of the silicon crowd: LLM ensemble prediction capabilities
  rival human crowd accuracy}.
\newblock \emph{Science Advances}, 2024.
\newblock \doi{10.1126/sciadv.adp1528}.

\bibitem[Schoenegger et~al.(2025)Schoenegger, Park, Karger, Trott, and
  Tetlock]{schoenegger2024ai}
Philipp Schoenegger, Peter~S. Park, Ezra Karger, Sean Trott, and Philip~E.
  Tetlock.
\newblock {AI-Augmented Predictions: LLM Assistants Improve Human Forecasting
  Accuracy}.
\newblock \emph{ACM Transactions on Interactive Intelligent Systems},
  15\penalty0 (1):\penalty0 1--25, 2025.
\newblock \doi{https://doi.org/10.1145/3707649}.

\bibitem[Skalse et~al.(2022)Skalse, Howe, Krasheninnikov, and
  Krueger]{skalse2022defining}
Joar Skalse, Nikolaus Howe, Dmitrii Krasheninnikov, and David Krueger.
\newblock {Defining and Characterizing Reward Hacking}, 2022.

\bibitem[Spooner et~al.(2018)Spooner, Fearnley, Savani, and
  Koukorinis]{spooner2018market}
Thomas Spooner, John Fearnley, Rahul Savani, and Andreas Koukorinis.
\newblock {Market Making via Reinforcement Learning}.
\newblock arXiv preprint arXiv:1804.04216, 2018.
\newblock URL \url{https://arxiv.org/abs/1804.04216}.

\bibitem[Troy et~al.(2026)Troy, Shields, Bradwell, and
  McCrory]{anthropic2026deal}
Kevin~K. Troy, Dylan Shields, Keir Bradwell, and Peter McCrory.
\newblock Project deal: Our {Claude}-run marketplace experiment.
\newblock Anthropic research report, 2026.
\newblock URL \url{https://www.anthropic.com/features/project-deal}.

\bibitem[Vaccaro et~al.(2024)Vaccaro, Almaatouq, and
  Malone]{vaccaro2024combinations}
Michelle Vaccaro, Abdullah Almaatouq, and Thomas~W. Malone.
\newblock {When combinations of humans and AI are useful: A systematic review
  and meta-analysis}.
\newblock \emph{Nature Human Behaviour}, 8\penalty0 (12):\penalty0 2293--2303,
  2024.
\newblock \doi{https://doi.org/10.1038/s41562-024-02024-1}.

\bibitem[Wang et~al.(2025)Wang, Wang, Yang, Patel, Zhao, Mo, Peng, Qian, Chen,
  Gutiérrez-Basulto, et~al.]{wang2025finauditing}
Yan Wang, Keyi Wang, Shanshan Yang, Jaisal Patel, Jeff Zhao, Fengran Mo,
  Xueqing Peng, Lingfei Qian, Yankai Chen, Víctor Gutiérrez-Basulto, et~al.
\newblock {FinAuditing}: A financial taxonomy-structured multi-document
  benchmark for evaluating {LLMs}, 2025.

\bibitem[Wang et~al.(2026)Wang, Han, Qian, He, Peng, Feng, Xie, Zhang, Guo, Mo,
  et~al.]{wang2026convfinre}
Yan Wang, Yi~Han, Lingfei Qian, Yueru He, Xueqing Peng, Dongji Feng, Zhuohan
  Xie, Vincent~Jim Zhang, Rosie Guo, Fengran Mo, et~al.
\newblock Conv-{FinRe}: A conversational and longitudinal benchmark for
  utility-grounded financial recommendation, 2026.

\bibitem[Wolfers and Zitzewitz(2004)]{wolfers2004prediction}
Justin Wolfers and Eric Zitzewitz.
\newblock {Prediction Markets}.
\newblock \emph{Journal of Economic Perspectives}, 2004.
\newblock \doi{10.1257/0895330041371321}.

\bibitem[Xiao et~al.(2025)Xiao, Jiang, Qian, Chen, He, Xu, Jiang, Li, Weng,
  Peng, et~al.]{xiao2025rafts}
Mengxi Xiao, Zihao Jiang, Lingfei Qian, Zhengyu Chen, Yueru He, Yijing Xu,
  Yuecheng Jiang, Dong Li, Ruey-Ling Weng, Min Peng, et~al.
\newblock Retrieval-augmented large language models for financial time series
  forecasting, 2025.

\bibitem[Xiao et~al.(2024)Xiao, Sun, Luo, and Wang]{xiao2024tradingagents}
Yijia Xiao, Edward Sun, Di~Luo, and Wei Wang.
\newblock {TradingAgents}: Multi-agents {LLM} financial trading framework.
\newblock \emph{arXiv preprint arXiv:2412.20138}, 2024.
\newblock \doi{10.48550/arXiv.2412.20138}.

\bibitem[Xie et~al.(2023)Xie, Han, Zhang, Lai, Peng, Lopez-Lira, and
  Huang]{xie2023pixiu}
Qianqian Xie, Weiguang Han, Xiao Zhang, Yanzhao Lai, Min Peng, Alejandro
  Lopez-Lira, and Jimin Huang.
\newblock {PIXIU: A Large Language Model, Instruction Data and Evaluation
  Benchmark for Finance}, 2023.

\bibitem[Xie et~al.(2024)Xie, Han, Chen, Xiang, Zhang, He, Xiao, Li, Dai, and
  Feng]{xie2024finben}
Qianqian Xie, Weiguang Han, Zhengyu Chen, Ruoyu Xiang, Xiao Zhang, Yueru He,
  Mengxi Xiao, Dong Li, Yongfu Dai, and Duanyu Feng.
\newblock {FinBen: A Holistic Financial Benchmark for Large Language Models}.
\newblock In \emph{Advances in Neural Information Processing Systems 37}, 2024.
\newblock \doi{10.52202/079017-3033}.

\bibitem[Xie et~al.(2025)Xie, Orel, Thareja, Sahnan, Madmoun, Zhang, Banerjee,
  Georgiev, Peng, Qian, et~al.]{xie2025finchain}
Zhuohan Xie, Daniil Orel, Rushil Thareja, Dhruv Sahnan, Hachem Madmoun, Fan
  Zhang, Debopriyo Banerjee, Georgi Georgiev, Xueqing Peng, Lingfei Qian,
  et~al.
\newblock {FinChain}: A symbolic benchmark for verifiable chain-of-thought
  financial reasoning, 2025.

\bibitem[Xiong et~al.(2025)Xiong, Deng, Wang, Cao, Li, Yu, Peng, Lin, Smith,
  Liu, et~al.]{xiong2025flagtrader}
Guojun Xiong, Zhiyang Deng, Keyi Wang, Yupeng Cao, Haohang Li, Yangyang Yu,
  Xueqing Peng, Mingquan Lin, Kaleb~E Smith, Xiao-Yang Liu, et~al.
\newblock {FLAG-Trader}: Fusion {LLM-Agent} with gradient-based reinforcement
  learning for financial trading, 2025.

\bibitem[Yu et~al.(2024{\natexlab{a}})Yu, Li, Chen, Jiang, Li, Zhang, Liu,
  Suchow, and Khashanah]{yu2024finmem}
Yangyang Yu, Haohang Li, Zhi Chen, Yuechen Jiang, Yang Li, Denghui Zhang, Rong
  Liu, Jordan~W. Suchow, and Khaldoun Khashanah.
\newblock {FinMem}: A performance-enhanced {LLM} trading agent with layered
  memory and character design.
\newblock In \emph{Proceedings of the AAAI Symposium Series}, volume~3, pages
  595--597, 2024{\natexlab{a}}.
\newblock \doi{10.1609/aaaiss.v3i1.31290}.

\bibitem[Yu et~al.(2024{\natexlab{b}})Yu, Yao, Li, Deng, Cao, Chen, Suchow,
  Liu, Cui, Xu, et~al.]{yu2024fincon}
Yangyang Yu, Zhiyuan Yao, Haohang Li, Zhiyang Deng, Yupeng Cao, Zhi Chen,
  Jordan~W. Suchow, Rong Liu, Zhenyu Cui, Zhaozhuo Xu, et~al.
\newblock {FinCon}: A synthesized {LLM} multi-agent system with conceptual
  verbal reinforcement for enhanced financial decision making,
  2024{\natexlab{b}}.

\bibitem[Zhang et~al.(2024{\natexlab{a}})Zhang, Zhao, Xia, Sun, Sun, Qin, Li,
  Zhao, Zhao, Cai, Zheng, Wang, and An]{zhang2024finagent}
Wentao Zhang, Lingxuan Zhao, Haochong Xia, Shuo Sun, Jiaze Sun, Molei Qin,
  Xinyi Li, Yuqing Zhao, Yilei Zhao, Xinyu Cai, Longtao Zheng, Xinrun Wang, and
  Bo~An.
\newblock A multimodal foundation agent for financial trading: Tool-augmented,
  diversified, and generalist.
\newblock In \emph{Proceedings of the 30th ACM SIGKDD Conference on Knowledge
  Discovery and Data Mining}, 2024{\natexlab{a}}.
\newblock \doi{10.1145/3637528.3671801}.

\bibitem[Zhang et~al.(2024{\natexlab{b}})Zhang, Xiang, Yuan, Feng, Han,
  Lopez-Lira, Liu, Ananiadou, Peng, Huang, et~al.]{zhang2024dolares}
Xiao Zhang, Ruoyu Xiang, Chenhan Yuan, Duanyu Feng, Weiguang Han, Alejandro
  Lopez-Lira, Xiao-Yang Liu, Sophia Ananiadou, Min Peng, Jimin Huang, et~al.
\newblock Dólares or dollars? unraveling the bilingual prowess of financial
  {LLMs} between {Spanish} and {English}, 2024{\natexlab{b}}.

\bibitem[Zheng et~al.(2022)Zheng, Trott, Srinivasa, Parkes, and
  Socher]{zheng2022economist}
Stephan Zheng, Alexander Trott, Sunil Srinivasa, David~C. Parkes, and Richard
  Socher.
\newblock The {AI} economist: Taxation policy design via two-level deep
  multiagent reinforcement learning.
\newblock \emph{Science Advances}, 8\penalty0 (18):\penalty0 eabk2607, 2022.
\newblock \doi{10.1126/sciadv.abk2607}.

\end{thebibliography}

\appendix
\section{Pre-Registered Simulation of Joint Design}
\label{app:sim}

The simulation illustrates the evaluation proposed in
Section~\ref{sec:aggregation}: the same budget, the same evidence, and the same
participants, with only the design of the market's functions varied. The
hypotheses, metrics, and comparisons were written down before any run
(\texttt{docs/sim\_prereg\_zh.md}); all are reported, including those that
failed. Code, beliefs, and figures are in the supplementary material
(\texttt{sim/}).

\subsection{Environment}

\paragraph{Claims and evidence.} Each world has 60 candidate contracts of the
form ``will dataset $X$ be released by day 12?''. Each has a prior drawn from
$\mathrm{Beta}(2,2)$, a true outcome drawn from it, and an ambiguity level
that makes settlement harder. Independent sources arrive at random days
(Poisson, mean 3 per contract) with signed strength; each found source may be
re-reported by aggregators (Poisson, mean 2), which adds no information.

\paragraph{Functions and design variables.} \emph{Information}: a monitor
spends effort $m$ per contract and finds each source with probability
$1-e^{-m}$; it is paid either per posted report, which makes posting re-reports
profitable, or only for new independent evidence. \emph{Liquidity}: an LMSR
market maker with parameter $b$ and worst-case subsidy $b\ln 2$.
\emph{Settlement}: review effort $r$ per contract; the chance of
mis-settlement is $a_i \exp(-(r_{\mathrm{eff}}/\rho + \gamma E_i))$, where
$a_i$ is ambiguity, $E_i$ the independent evidence found, and
$r_{\mathrm{eff}}$ the effort left after processing re-reports. $\gamma$ is the
evidence--settlement coupling we expected to drive the gains of joint design.
\emph{Trading rule}: continuous trading, in which agents act before slow
traders within each step, or uniform-price batch clearing once per step.
Every listed contract costs $m + b\ln 2 + r$ from a shared budget of 400, and
a contract is listed only if its expected mis-settlement rate is at most 5\%.

\paragraph{Participants.} Ten principals delegate to strong agents, which
discount re-reports and combine public evidence with the principal's private
signal by Bayes' rule; ten delegate to weak agents, which count re-reports as
independent and read evidence with noise; ten slow traders hold only
private signals and trade in 30\% of steps. Each participant holds a private
signal on a contract with probability 0.3.

\paragraph{Designs and objective.} The \emph{fixed} design uses hand-set
defaults. In the \emph{separate} design each function optimizes its own
objective: the monitor its reports (a learned monitoring policy chooses
$m=3$ and posts re-reports), the market maker its loss (smallest $b$), the
reviewer the worst case without evidence, and the exchange its volume
(continuous trading). The \emph{joint} design maximizes the market objective
$J$, the sum over listed contracts of the Brier improvement of the final
price over the prior, minus 0.5 per mis-settled contract, by grid search
subject to the slow traders' average profit being at least $-2$. All tuning
uses worlds 0--9; all reported numbers use worlds 100--139, paired across
designs. Sensitivity runs vary re-report volume (0 or 4) and budget (250 or
600).

\subsection{Pre-registered hypotheses and outcomes}

\begin{table}[!htb]
\centering
\small
\begin{tabular}{@{}p{0.29\columnwidth}p{0.62\columnwidth}@{}}
\toprule
Hypothesis & Outcome (40 held-out worlds)\\
\midrule
H1 Joint design wins, more so as $\gamma$ grows & Partly. $+2.07 \pm 0.22$, 37 of 40 worlds; the gain of jointly tuning monitoring and review falls from $+3.0$ to $-0.3$ as $\gamma$ rises from 0 to 1.\\
H2 Per-report pay induces re-reports & Yes, largely by construction. Paying for new evidence cuts mis-settlement from 2.23 to 0.80; $J$ moves from 2.62 to 2.71.\\
H3 Batch clearing protects slow traders & Yes. Slow traders' profit $-3.4 \to +9.2$, $J$ $2.53 \to 4.69$; no effect without slow traders.\\
H4 Decompose the shortfall & Order-dependent: information $+0.09$ first, $+0.33$ last; liquidity $+0.37$ first, $-0.03$ last; 52\% of the gap remains with participants.\\
H5 Weak-agent principals lose, unnoticed & Direction yes; gap/s.d. 0.85 (fixed), 1.39 (joint).\\
\bottomrule
\end{tabular}
\caption{Pre-registered hypotheses and what the stylized simulation found.}
\label{tab:prereg}
\end{table}

\subsection{Language-model agents}

\paragraph{Evidence as news.} For six held-out worlds (180 contracts), each
found source was written as a dated news item (an official changelog, a
maintainer's statement) whose wording reflects its strength. Each re-report
restates the same fact in different words, and half name an outlet; none
repeats the identifier in the original item, so a careful reader can
recognize but not trivially match it. This gives 481 independent items and
992 re-reports. The \emph{clean} condition shows independent items only,
corresponding to paying for new evidence; the \emph{echo} condition adds the
re-reports, corresponding to paying per report.

\paragraph{Elicitation and replay.} Claude Haiku~4.5 (weak reader) and Claude
Sonnet (strong reader) each read every contract under both conditions, 30
contracts per session, and reported a probability for each of the 12 days.
Instructions were neutral and did not mention re-reports. This produced 24
sessions and 720 belief paths. As in Project Deal and Project Swap
\citep{anthropic2026deal,anthropic2026swap}, beliefs were elicited once and
replayed across all market designs, so differences between designs cannot
come from sampling the models again. Principals' private signals and slow
traders are as in the stylized model; only reading the news is done by the
models.

\begin{figure*}[t]
\centering
\includegraphics[width=\textwidth]{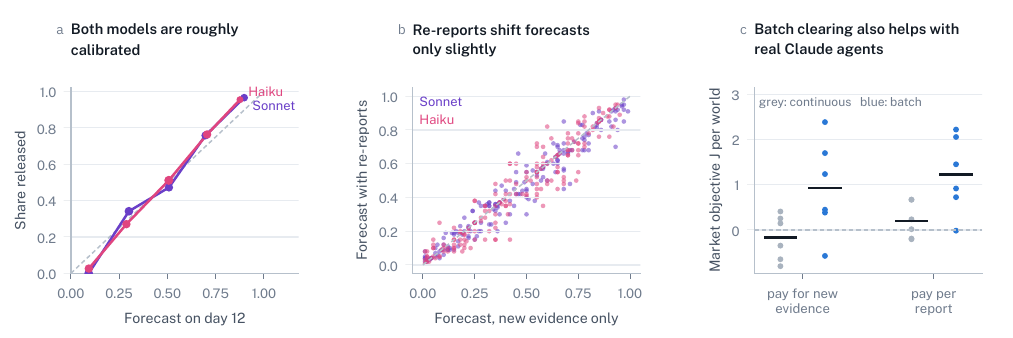}
\caption{\textbf{Claude agents, six held-out worlds, 180 contracts.} (a)
Share of contracts released by day-12 forecast bin. (b) Day-12 forecast with
and without re-reports, per contract and model. (c) Market objective per world
under continuous (grey) and batch (blue) trading; bars are means.}
\label{fig:llm}
\end{figure*}

\paragraph{Results.} Both models are roughly calibrated
(Figure~\ref{fig:llm}a). Re-reports change day-12 Brier scores only slightly
(Sonnet 0.139 to 0.145, Haiku 0.151 to 0.150; Figure~\ref{fig:llm}b), so the
stylized assumption that weak agents double count does not hold for these
models; why the gap shrinks under per-report pay
(re-report timing, trading on belief paths) is not isolated here. Batch clearing raises $J$ in both conditions
($+1.03 \pm 0.38$ per report, five of six worlds; $+1.09 \pm 0.27$ for new
evidence, six of six; Figure~\ref{fig:llm}c) but slightly lowers slow traders'
profit ($-0.72 \pm 0.42$ and $-0.83 \pm 0.63$): with the models' more
moderate beliefs, agents no longer trade far ahead of slow traders. The gap
between principals represented by Sonnet and by Haiku is $12.5 \pm 2.6$
(continuous) and $11.5 \pm 3.5$ (batch) when the monitor is paid for new
evidence, positive in all six worlds, and $2.9 \pm 3.9$ and $1.4 \pm 4.6$ when
it is paid per report.

\subsection{Limitations}

Participants other than the news readers are stylized, and every parameter is
set by us, so the numbers indicate the direction of mechanisms rather than
their size in any market. The benefit of batch clearing depends on agents
acting before slow traders in continuous trading. That the separate design's
exchange chooses continuous trading is an assumption about its objective.
Six language-model worlds give large standard errors and should be read as a
pilot. The news is generated from templates and is more regular than real
reporting, and a model reading 30 contracts in one session may carry
impressions between them.

\end{document}